# Electrical Control of Plasmon Resonance with Graphene

Jonghwan Kim[1], Hyungmok Son[1], David J. Cho[1,2], Baisong Geng[1], Will Regan[1,2], Sufei Shi[1], Kwanpyo Kim[1,2], Alex Zettl[1,2], Yuen-Ron Shen[1,2], and Feng Wang[1,2]

[1] Department of Physics, University of California at Berkeley, Berkeley, CA 94720, USA.

[2] Materials Science Division, Lawrence Berkeley National Laboratory, Berkeley, CA 94720, USA.

**Surface plasmon, with its unique capability to concentrate light into sub-wavelength volume, has enabled great advances in photon science, ranging from nano-antenna[1] and single-molecule Raman scattering[2] to plasmonic waveguide[3] and metamaterials[4,5]. In many applications it is desirable to control the surface plasmon resonance in situ with electric field. Graphene, with its unique tunable optical properties[6,7], provides an ideal material to integrate with nanometallic structures for realizing such control. Here we demonstrate effective modulation of the plasmon resonance in a model system composed of hybrid graphene-gold nanorod structure. Upon electrical gating the strong optical transitions in graphene can be switched on and off, which leads to significant modulation of both the resonance frequency and quality factor of plasmon resonance in gold nanorods. Hybrid graphene-nanometallic structures, as exemplified by this combination of graphene and gold nanorod, provide a general and powerful way for electrical control of plasmon resonances. It holds promise for novel active optical devices and plasmonic circuits at the deep sub-wavelength scale.**

Surface plasmon resonance in nanoscale metal structures has attracted tremendous interest due to its unique capability to concentrate light into deep subwavelength scale. Recent advances in plasmonics have enabled a wide spectrum of applications, such as single-molecule Raman scattering[2], sub-diffraction-limit imaging[8], and optical cloaking[4,5]. All these applications rely on our capability to control the surface plasmon resonance through structural design, as embodied in shape-controlled metal nanoparticles[9], nano-antennas[1], and plasmonic metamaterials[4,5]. New opportunities can emerge if we can develop new capability in surface plasmon control, such as high-speed, in-situ modulation of plasmon resonances in an existing nanometallic structure.

In general the surface plasmon resonance can be modulated by varying the dielectric environment around a nanometallic structure[10]. In previous studies this has been achieved by modifying the molecules surrounding the metal, such as through molecular adsorption[11], chemical reaction[12], or liquid crystal alignment[13]. These methods, however, are intrinsically very slow and they are often irreversible and difficult to control. A more attractive approach is to use hybrid semiconductor-metal nanostructures, where the semiconductor electronic transitions (, and therefore its dielectric constant,) can be varied through electrostatic gating[14,15]. This approach, however, has so far been hampered by a lack of suitable semiconductor materials that have large gate-induced dielectric constant change and can be incorporated at the hot spots of plasmon excitation. Graphene, a novel zero-bandgap semiconductor[16,17], satisfies these stringent requirements and can be ideal for electrical control of plasmon resonances: optical absorption of graphene is among the strongest in all materials, and remarkably, it can be conveniently switched on and off across a broad spectral range with electrical gating[6,7]. In addition, graphene can be readily integrated into the nanometer-sized hot spot in the nanometallic structure due to its single atom thickness and excellent compatibility with nanofabrication.

Here we demonstrate in-situ electrical control of plasmon resonance using the hybrid graphene-gold nanorod structure as a model system. We show that electrical gating of graphene can significantly modulate all aspects of the plasmon resonance, including a 20 meV shift of resonance frequency, a 30% increase in quality factor, and a 30% increase in resonance scattering intensity. The plasmon resonance frequency shift and the resonance quality factor increase can be attributed to, respectively, changes in the real ($\varepsilon_g'$) and imaginary ($\varepsilon_g''$) part of the graphene dielectric constant when optical transitions in graphene are Pauli blocked upon electrical gating. This mechanism is generally applicable in hybrid graphene-nanometallic structures for electrical control of surface plasmon. Plasmon resonances in such hybrid structures, unlike the pure graphene plasmon limited to terahertz frequencies[18,19], can cover any spectral range through suitable design of the nanometallic structures. And the electrical control can potentially be realized at ultrahigh speed given the excellent electrical transport properties of graphene, as demonstrated in ultrafast graphene photodetectors[20] and modulators[21]. It will open up exciting new opportunity for in-situ, high-speed control of light at deep subwavelength scale in nanoplasmonic devices.

Fig. 1a illustrates our device configuration, where graphene was placed on top of gold nanorods. Fig. 1b shows a typical high-resolution scanning electron micrograph (SEM) of the single gold nanorod covered by graphene in our devices. We used top electrolyte gating with ionic liquid to control optical transitions in graphene[22]. The consequent changes in plasmon resonance of individual graphene-nanorod hybrid structures were probed with dark-field Rayleigh scattering spectroscopy.

Fig. 2a and 2b display two representative single-particle Rayleigh scattering spectra of an ungated graphene-nanorod hybrid structure and a bare gold nanorod, respectively. Both scattering spectra exhibit prominent plasmon resonances at 0.86 eV (i.e. at the telecom wavelength of 1.5 µm). However, the plasmon resonance of the graphene-nanorod hybrid structure is much broader, with a full-width-half-maximum (FWHM) at 93 meV (compared to 70 meV in a bare gold nanorod). It demonstrates that the monolayer graphene right above the gold nanorod can strongly affect the plasmon resonance. Similar changes in plasmon resonance have been observed by depositing graphene on top a metamaterial[23]. The increased plasmon resonance width can be understood as a consequence of the strong interband optical absorption of pristine graphene at 0.86 eV (Fig. 2c)[24,25], which provides an efficient dissipation channel and increases the damping rate of the surface plasmon oscillation.

To modulate the plasmon resonance, one can simply eliminate the energy dissipation in graphene by switching off its interband optical transitions. This can be achieved through electrostatic gating: gated graphene has a shifted Fermi energy $|E_F|$, and optical transitions with energy less than $2|E_F|$ become forbidden due to empty initial states (or filled final states) for hole (or electron) doping (Fig. 2d)[6,7]. We can determine this gate-induced change in graphene absorption using infrared reflection spectroscopy. Fig. 3a displays the gate-induced reflectivity change δR/R at the plasmon resonance energy 0.86 eV in an area with only graphene. It shows a step-function-like decrease in reflectivity, corresponding to reduced graphene absorption[25], at large hole doping (with $V_g$ lower than − 0.1 V). This gate-dependent reflectivity curve indicates that our as-prepared graphene under ionic liquid is strongly hole doped, and $2|E_F|$ reaches the probe photon energy of 0.86 eV at $V_g$ = − 0.1 V. Similar curves obtained at different probe photon energies allow us to determine the $2|E_F|$ value of graphene at different applied gate voltages.

Rayleigh scattering intensity from an individual graphene-nanorod hybrid structure as a function of the photon energy and gate voltage is displayed in a two-dimensional (2D) color plot in Fig. 3b. Fig. 3c shows four line cuts of the 2D plot for Rayleigh scattering spectra at $V_g$ = 0.5, -0.1, -0.9, and -1.5V, which clearly demonstrate the capability to modulate surface plasmon resonance through electrical gating. The detailed dependence of plasmon resonance energy ($E_R$), width ($\Gamma_R$), and peak intensity ($I_P$) on graphene $2|E_F|$ are shown in Fig. 4a-c (symbols). The resonance width in Fig. 4b displays a step-like decrease, corresponding to an increase in quality factor, with increasing $2|E_F|$. This is a direct consequence of blocked graphene optical absorption in highly doped graphene, which leads to a reduced $\varepsilon_g^{"}$ in graphene and lower loss. The increased quality factor naturally leads to a higher scattering intensity at the plasmon resonance, as shown in Fig. 4c. The plasmon resonance frequency exhibits an unusual behavior (Fig. 4a): it shifts to lower energy, and then to higher energy, with increased graphene doping. This behavior can be accounted for by gated-induced change in the real part of dielectric constant ($\varepsilon_g^{'}$) of graphene, as we describe below.

Gate-dependent complex dielectric constant of graphene has been established previously[25-27]. The imaginary part $\varepsilon_g^{"}$ is characterized by a constant absorption of $\frac{\pi e^2}{\hbar c}$ above $2|E_F|$ from interband transitions and Drude absorption from free carriers (i.e. intraband transitions). The real part $\varepsilon_g^{'}$ can be obtained from $\varepsilon_g^{"}$ using the Kramer-Kronig relation. Specifically, energy-dependent $\varepsilon_g^{'}$ and $\varepsilon_g^{"}$ have the form[25-27]

$$\varepsilon"_g(E) = \frac{e^2}{4E\varepsilon_0 d}\left[1 + \frac{1}{\pi}\left\{\tan^{-1}\frac{E-2|E_F|}{\Gamma} - \tan^{-1}\frac{E+2|E_F|}{\Gamma}\right\}\right] + \frac{e^2}{\pi\tau E\varepsilon_0 d}\frac{|E_F|}{E^2+(1/\tau)^2} \text{ and}$$

$$\varepsilon'_g(E) = 1 + \frac{e^2}{8\pi E\varepsilon_0 d}\ln\frac{(E+2|E_F|)^2+\Gamma^2}{(E-2|E_F|)^2+\Gamma^2} - \frac{e^2}{\pi\varepsilon_0 d}\frac{|E_F|}{E^2+(1/\tau)^2}$$

Here $d$ is the thickness of graphene. The interband transition broadening $\Gamma$ is estimated to be 110 meV from the graphene reflection spectrum. The free carrier scattering rate $1/\tau$ can be set to zero because it has little effect on the dielectric constants, $\varepsilon_g^{'}(E_R)$ and $\varepsilon_g^{"}(E_R)$, at the plasmon resonance energy $E_R$. The equations show that $\varepsilon_g^{"}(E_R)$ experiences a step like decrease when $2|E_F|$ is larger than $E_R$ and blocks the relevant interband transitions. However, $\varepsilon_g^{'}(E_R)$ has a maximum at $2|E_F| = E_R$. This is because all optical transitions below $E_R$ contribute a negative susceptibility, and transitions above $E_R$ contribute a positive susceptibility to $\varepsilon_g'(E_R)$.

The changes in $\varepsilon_g'$ and $\varepsilon_g''$ will modify, respectively, the resonance energy and width of the plasmon excitation. For simplicity we will treat the effects of graphene on the plasmon resonance using perturbation theory, where the frequency shift and width increase of the plasmon resonance are proportional to the gate-dependent $\varepsilon_g'$ and $\varepsilon_g''$ of graphene, i.e., $E_R = E_R^0 + \alpha \varepsilon_g'$ and $\Gamma_R = \Gamma_R^0 + \beta \varepsilon_g''$. Here $E_R^0$ and $\Gamma_R^0$ are the plasmon resonance frequency and width for bare gold nanorod, and $\alpha$ and $\beta$ are two constant prefactors. Solid lines in Fig. 4a and 4b show a fitting to the experimental data with the initial resonance energy $E_R^0 = 862.4\ meV$, resonance width $\Gamma_R^0 = 72.8\ meV$, and the overall scaling factors $\alpha = -2.124$, $\beta = 0.7158$. (For best fitting we have also slightly shifted $|E_F|$ by 40 meV compared to the values we determined through optical spectroscopy at the region with only graphene. This shift is presumably due to slightly different carrier doping at the hot spot right next to gold.) Our simple model reproduces nicely the significant gate-induced decrease of plasmon resonance width (i.e. higher quality factor), as well as the red and blue shifts of the plasmon resonance energy. The increased scattering intensity is arises naturally from the increased plasmon resonance quality factor.

In conclusion, we have demonstrated effective control of gold-nanorod plasmon resonance through electrostatic gating of graphene. This approach can be easily generalized for electrical control of plasmon resonances in other hybrid graphene-metal nanostructures, and can be readily improved in several key aspects. For example, stronger modulation can be achieved by utilizing sharper resonances of "dark" plasmon modes in metamaterials[28], or by using multilayer of graphene[19]. Higher speed operation can be realized using field effect configuration as that employed in large-bandwidth transistors[20] and waveguide modulators[21]. Such improvement will open up exciting new opportunities for high-speed in-situ control of light in deep subwavelength plasmonic structures. One can further envision to control graphene doping through optical excitation rather than electrical gating, which can enable ultrafast all-optical switching of plasmon excitations in such graphene-nanometallic structures.

**Methods**

Chemical synthesized gold nanorods were purchased from Nanopartz (part number: 30-HAR-1400). The gold nanorods have a mean diameter of 20 nm and length of 256 nm, with a plasmon resonance energy at 860 meV. Gold nanorods were deposited on a glass or $SiO_2$/Si substrate by spin coating at a spinning speed of 500 rpm. The substrate with gold nanorods was then immersed in acetone at 70 ℃ for 30 minutes to dissolve residual cetrimonium bromide (CTAB). On top of the nanorods we transferred a large-area graphene grown by chemical vapor deposition using the standard growth and transfer processes[29,30]. For electrostatic gating of graphene we used a top electrolyte gating with ionic liquid 1-ethyl-3-methylimidazolium bis(trifluoromethylsulfonyl)imide (EMI-TFSI)[22]. Fig. 1b displays a typical high-resolution scanning electron micrograph (SEM) of the gold nanorod covered by graphene on $SiO_2$/ Si substrate. Our SEM images show that the gold nanorods have a length distribution from 244 to 268 nm, and they form mostly well-separated individual rods in our devices. Transferred large area graphene draped nicely around the gold nanorods, as can be seen at the nanorod edges in the SEM image.

The electrically tunable plasmon resonance in the hybrid graphene-nanorod structure is probed using a dark-field Rayleigh scattering spectroscopy of individual gold nanorods. We used a supercontinuum laser as the broadband light source producing high brightness photons from 0.67 to 2.7 eV. The supercontinuum light is focused to excite gold nanorods in a microscopy setup. The Rayleigh scattering light from individual nanorods is collected using confocal microscopy in dark-field configuration and analyzed by a spectrometer equipped with an InGaAs array detector.

**Acknowledgements** We thank R. Segalman and B. Boudouris for providing the ionic liquid and thank X. Zhang for helpful discussion. This work was supported by the Department of Energy Early Career Award DE-SC0003949 (J.K., H.S., and F.W.) and by Office of Basic Energy Sciences, U.S. Department of Energy under Contract No. DE-AC02-05CH11231 (D.J.C., Y.R.S., and F.W.) and DE-AC03-76SF0098 (W.R. and A.Z.) for the Materials Science Division and the user facility of DOE Molecular Foundry (No. DE-AC02-05CH11231). F.W. also acknowledges the support from a David and Lucile Packard fellowship.

**Author contributions** F.W. designed the experiment; J.K., H.S. and D.J.C. carried out optical measurements; J.K., H.S., B.G., W.R., S.S., K.K., and A.Z. contributed to sample growth, fabrication and characterization. J.K. Y.R.S, and F.W. performed theoretical analysis. All authors discussed the results and wrote the paper together.

**Author information** The authors declare no competing financial interests. Correspondence and requests for materials should be addressed to F.W. (fengwang76@berkeley.edu).

**Figure 1: Graphene-gold nanorod hybrid structure**. **a**. Illustration of a typical device with graphene placed on top of gold nanorods. A top electrolyte gate with ionic liquid is used to control plasmon resonance through varying optical transitions in graphene. **b**. A typical high-resolution scanning electron micrograph (SEM) of the single gold nanorod covered by graphene. Individual gold nanorods are well-separated from each other in our devices, and graphene is observed to drape nicely over the nanorods (white arrow).

**Figure 2: Effect of graphene on the gold nanorod plasmon resonance. a**. Rayleigh scattering spectra of an as-prepared graphene-nanorod hybrid structure. **b**. Rayleigh scattering spectrum of a bare gold nanorod. Both scattering spectra exhibit prominent plasmon resonances at 0.86 eV (i.e. at the telecom wavelength of 1.5 µm), but the resonance in graphene-nanorod hybrid structure is significantly broader due to extra dissipation channel from graphene absorption. **c**. Illustration of strong interband optical transitions present at all energies in pristine graphene. They contribute to the plasmon dissipation at 0.86 eV. **d**. Illustration showing that gate-induced shift in Fermi energy ($E_F$) can block the interband transition in graphene and reduce optical dissipation at 0.86 eV.

**Figure 3: Electrical control of the plasmon resonance. a**. Gate-induced reflectivity change ($\delta R/R$) of graphene on the substrate probed at photon energy of 0.86 eV. It shows a step-function-like decrease in reflectivity, corresponding to a reduction of graphene absorption, due to blocked optical absorption at large hole doping (Fig. 2d). The threshold voltage at $V_g = -0.1$ V

is set by $2|E_F|$ reaching the probe photon energy of 0.86 eV. (It indicates that the as-prepared graphene under ionic liquid is strongly hole doped.) **b**. Scattering intensity (color scale, arbitrary units) is plotted as a function of the photon energy and gate voltage. **c**. Rayleigh scattering spectra of an individual graphene-nanorod hybrid structure at $V_g$ = 0.5, -0.1, -0.9, and -1.5 V, corresponding to the four horizontal (dashed) line cuts in **b**. Strong modulation of all aspects of the plasmon excitation, including the resonance frequency, quality factor, and scattering intensity, is achieved as electrostatic gating shifts the Fermi energy and modifies optical transitions in graphene.

**Figure 4: Comparison between experiment and theory.** Symbols in **a**, **b**, and **c** show, respectively, detailed data of the plasmon resonance energy ($E_R$), width ($\Gamma_R$), and peak scattering intensity ($I_P$) as a function of $2|E_F|$ in graphene. Our model (solid line) reproduces nicely the experimental data, where the changes in the plasmon resonance energy and width originate from gate-induced modification in the real ($\varepsilon_g'$) and imaginary ($\varepsilon_g''$) part of graphene dielectric constant, respectively. The red and then blue shift of plasmon resonance frequency in **a** is due to an increase and then decrease of $\varepsilon_g'$ in graphene at the plasmon resonance energy $E_R$ when optical transitions with increasingly higher energies are blocked. This is because interband transitions with energies lower than $E_R$ contributes a negative susceptibility, and transitions with energies higher than $E_R$ contributes a positive susceptibility to $\varepsilon_g'$ at $E_R$. The decreased resonance width at large $2|E_F|$ in **b** is a consequence of reduced $\varepsilon_g''$ and lower loss when optical transitions

at $E_R$ are blocked. This increased quality factor naturally leads to a higher scattering intensity at the plasmon resonance in **c**.

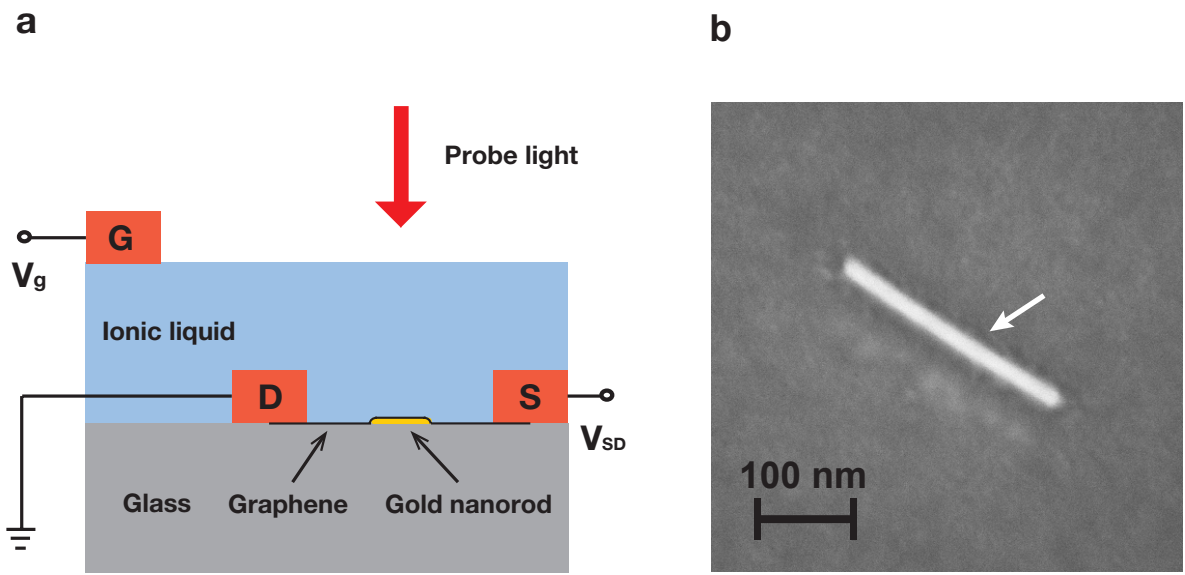

Figure 1

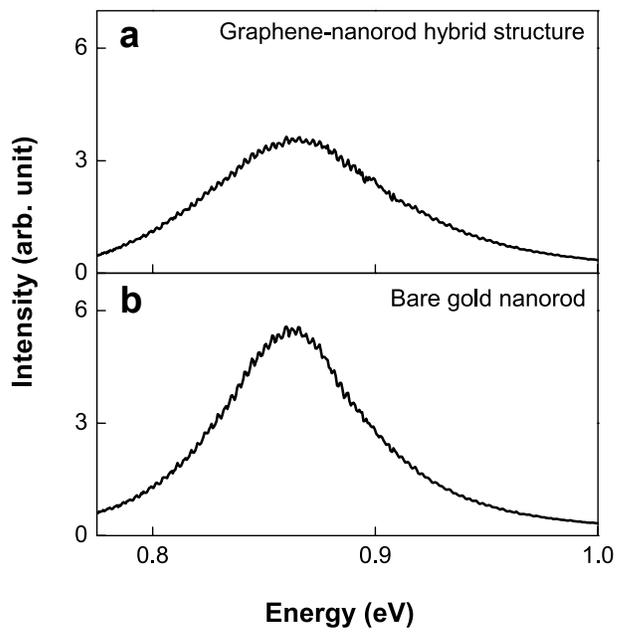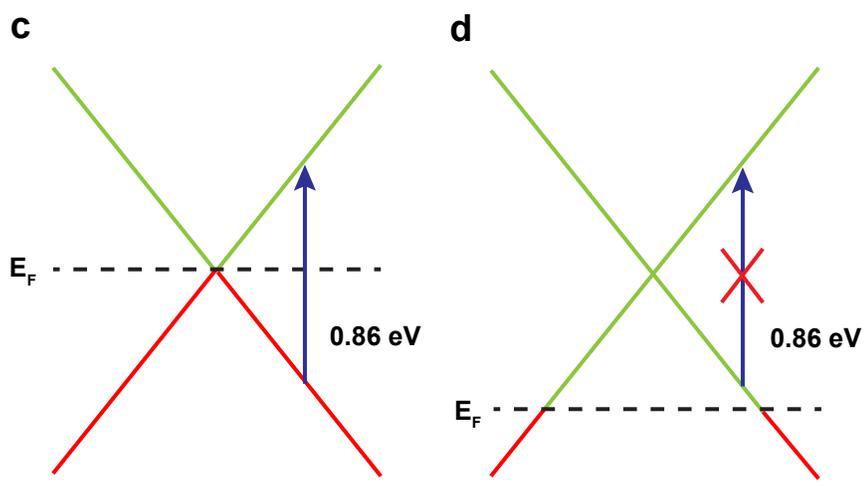

Figure 2

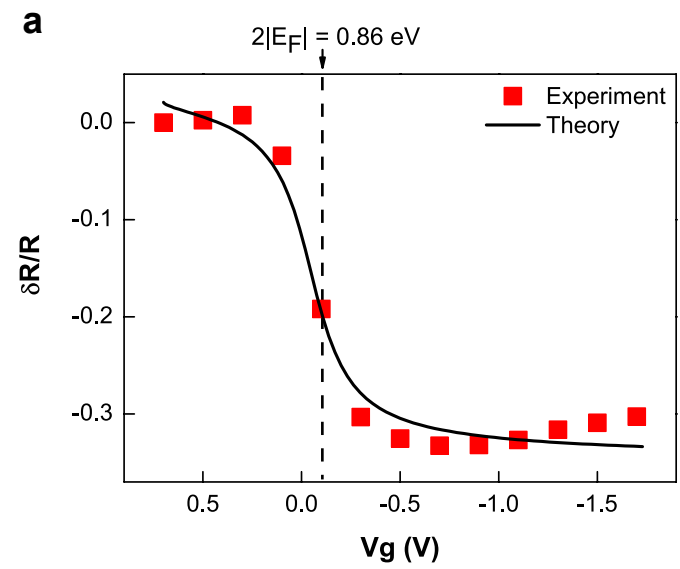

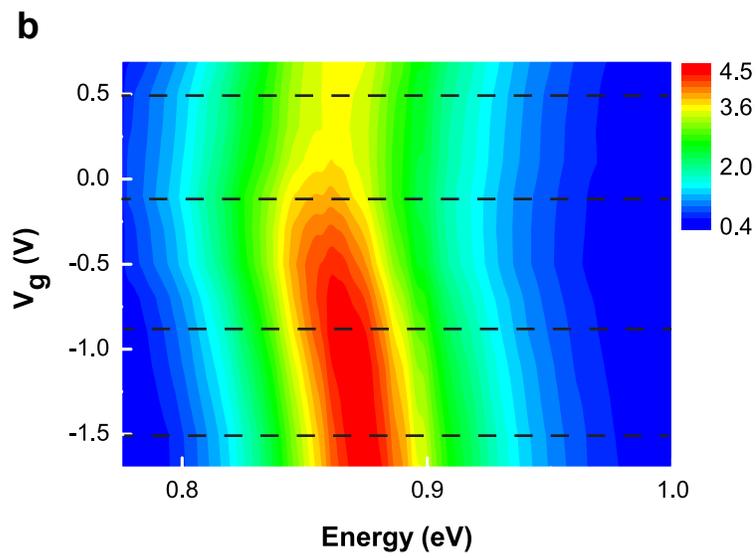
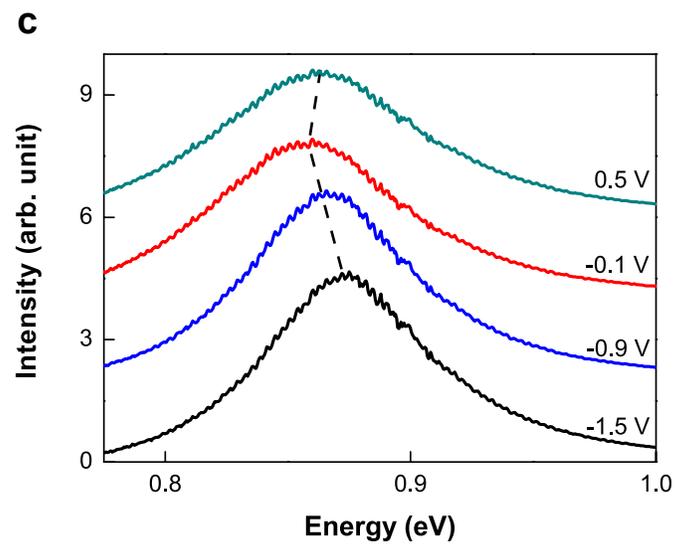

**Figure 3**

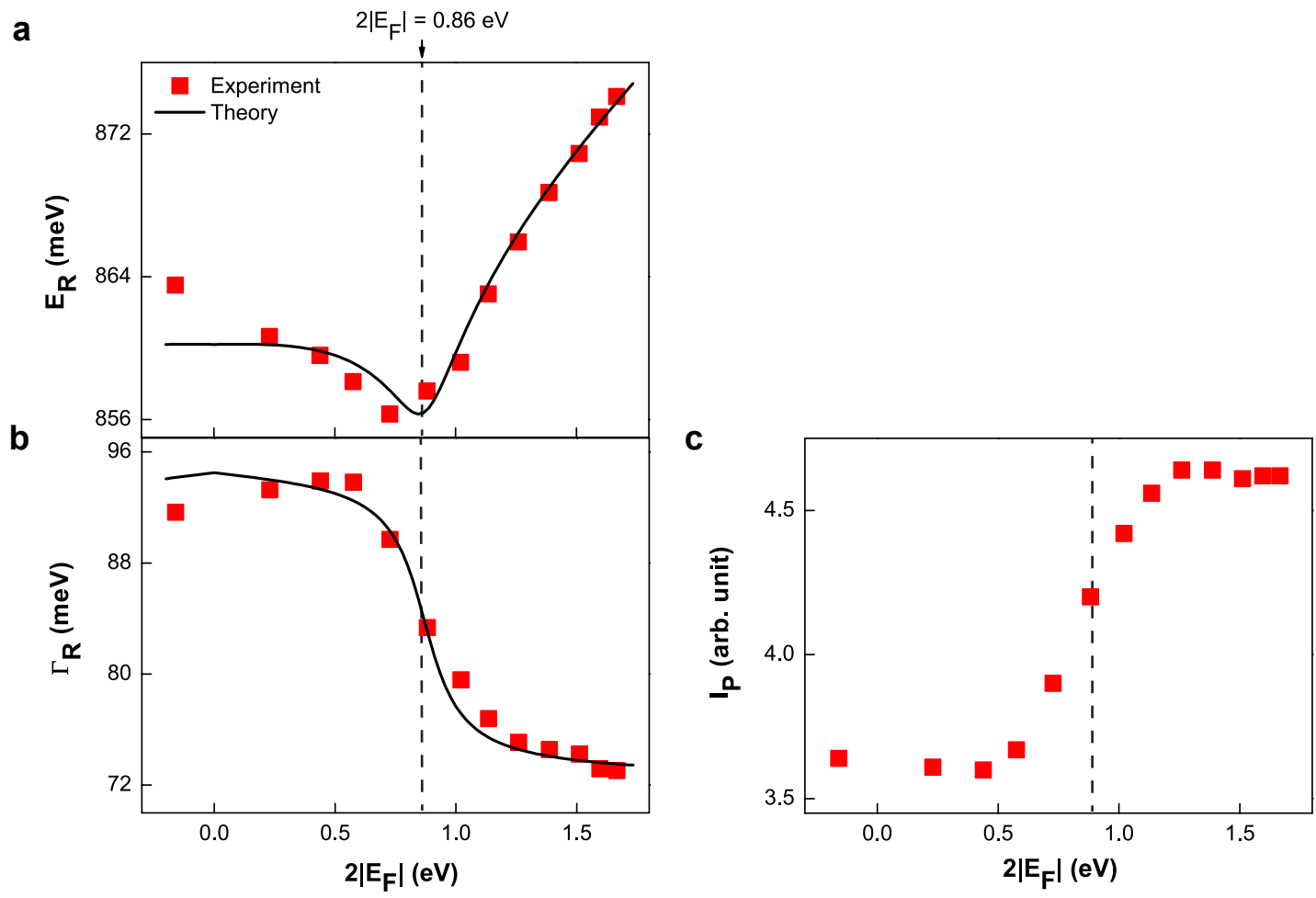

Figure 4